\documentclass[12pt]{iopart}
\usepackage{latexsym}
\usepackage{amssymb}
\usepackage{verbatim}
\usepackage{graphicx}
\usepackage{dcolumn}
\usepackage{bm}
\usepackage{setspace}
\usepackage{url}

\newcommand{\la}{\left\langle}
\newcommand{\ra}{\right\rangle}
\newcommand{\EPL}{\emph{Europhys.~Lett.~}}
\newcommand{\MP}{\emph{Mol.~Phys.~}}

\begin{document}

\noindent
In {\it Electrostatics of Soft and Disordered Matter}, D.~S.~Dean, J.~Dobnikar, 
A.~Naji, and R.~Podgornik, Eds., Proceedings of the CECAM Workshop 
``New Challenges in Electrostatics of Soft and Disordered Matter" 
(Pan Stanford, 2013)

\title[Coarse-Grained Modeling of Charged Colloids]
{Coarse-Grained Modeling of  Charged Colloidal Suspensions: 
From Poisson-Boltzmann Theory \\
to Effective Interactions}

\author{Alan R.~Denton}

\address{Department of Physics, North Dakota State University,
Fargo, North Dakota, U.S.A. \\[1ex]
{\tt alan.denton@ndsu.edu}}


\section{Introduction}\label{intro}
Electrostatic forces -- the strongest interparticle forces outside of
the nucleus -- account for the stability of matter over a broad range of
length scales, from atoms to macromolecules to the myriad materials
that surround us.  Colloidal particles (nanometers to microns in size)
and polymers can become charged in a polar solvent (e.g., water) through 
dissociation of counterions~\cite{evans99}.  Repulsive Coulomb interactions 
between ions can then stabilize a suspension or solution against aggregation 
due to ubiquitous van der Waals attractive interactions~\cite{israelachvili92}.  
Electrostatic interactions between ions largely govern the equilibrium thermodynamic 
and dynamical properties of charge-stabilized colloidal suspensions and 
polyelectrolyte solutions.  Controlling mechanical and thermal stability 
is essential to many applications -- from foods and pharmaceuticals
to filters and photonic materials.

Predicting the properties of such complex, multicomponent systems 
with accuracy sufficient to guide and interpret experiments requires 
realistic modeling of the interparticle interactions and collective 
behavior of many-particle systems.  While the fundamental interactions
are simple, the sheer number of particles and the broad ranges of length 
and time scales confront the modeler with significant computational 
challenges.  A general strategy for mitigating such challenges is to 
``coarse grain" or ``integrate out" the degrees of freedom of some components,
reducing the original multicomponent model to a simpler model of fewer 
components~{\cite{belloni00,hansen-lowen,likos01,zvelindovsky07}.
The trade-off for so reducing complexity is that the simpler model is 
governed by modified (effective) interparticle interactions.

This chapter is a ``how-to" manual of sorts for implementing coarse-graining 
methods to derive effective interactions.  For simplicity, we focus here on 
charge-stabilized colloidal suspensions.  The same basic concepts apply,
however, to a wide variety of soft (and hard) materials.
To set the stage, we begin by defining the primitive and one-component models 
and outlining the Poisson-Boltzmann (PB) theory of charged colloids.  After 
reviewing the well-known cell model implementation of PB theory, we turn to 
an alternative implementation, based on perturbation theory, and derive 
microion distributions around colloids and effective electrostatic interactions.
After some effort, a happy result emerges: the effective interactions remain 
relatively simple, at least in systems with monovalent microions.  This fortunate 
outcome provides the foundation for further theoretical and simulation modeling 
to explore and facilitate design of novel materials.  Finally, we peer over 
the horizon at the outlook for possible future research directions in the field.

\section{Primitive Model}\label{pmsec}
The primitive model of charged colloids and polyelectrolytes~\cite{evans99,israelachvili92}
idealizes the solvent as a homogeneous dielectric continuum of relative permittivity 
$\epsilon$.  Dispersed throughout the solvent are macroions and microions, modeled 
here as charged hard spheres of radius $a$ and valence $Z$ (charge $-Ze$) and point ions
of valence $z$.
In a closed suspension, all particles are confined to the same volume $V$.
In Donnan equilibrium, only the macroions are confined, while the microions
can exchange (e.g., across a semi-permeable membrane) with a salt reservoir, 
here assumed to be a 1:1 electrolyte of monovalent ions.  

\begin{figure}[h!]
\begin{center}
\includegraphics[width=0.34\columnwidth]{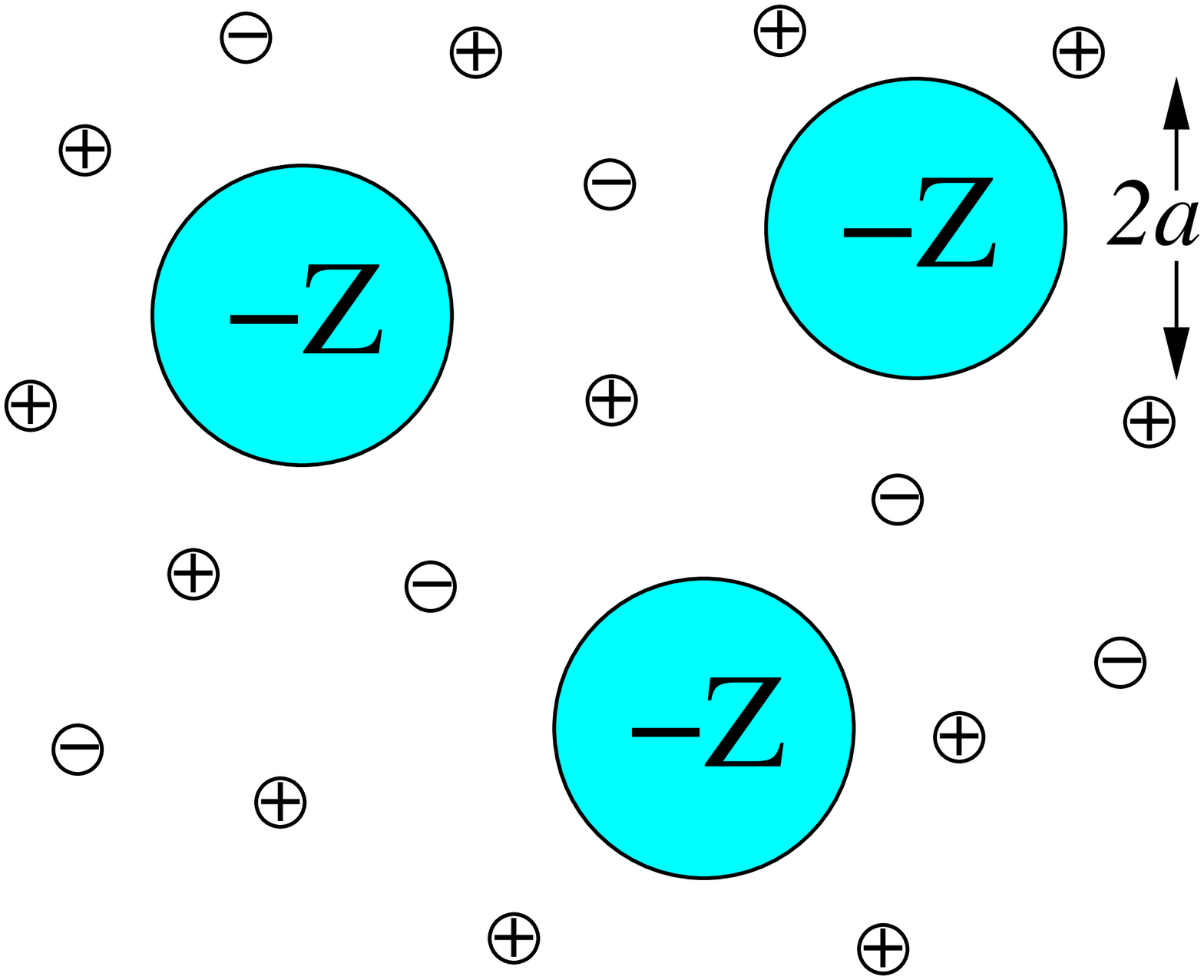}
\hspace*{1cm}
\includegraphics[width=0.49\columnwidth]{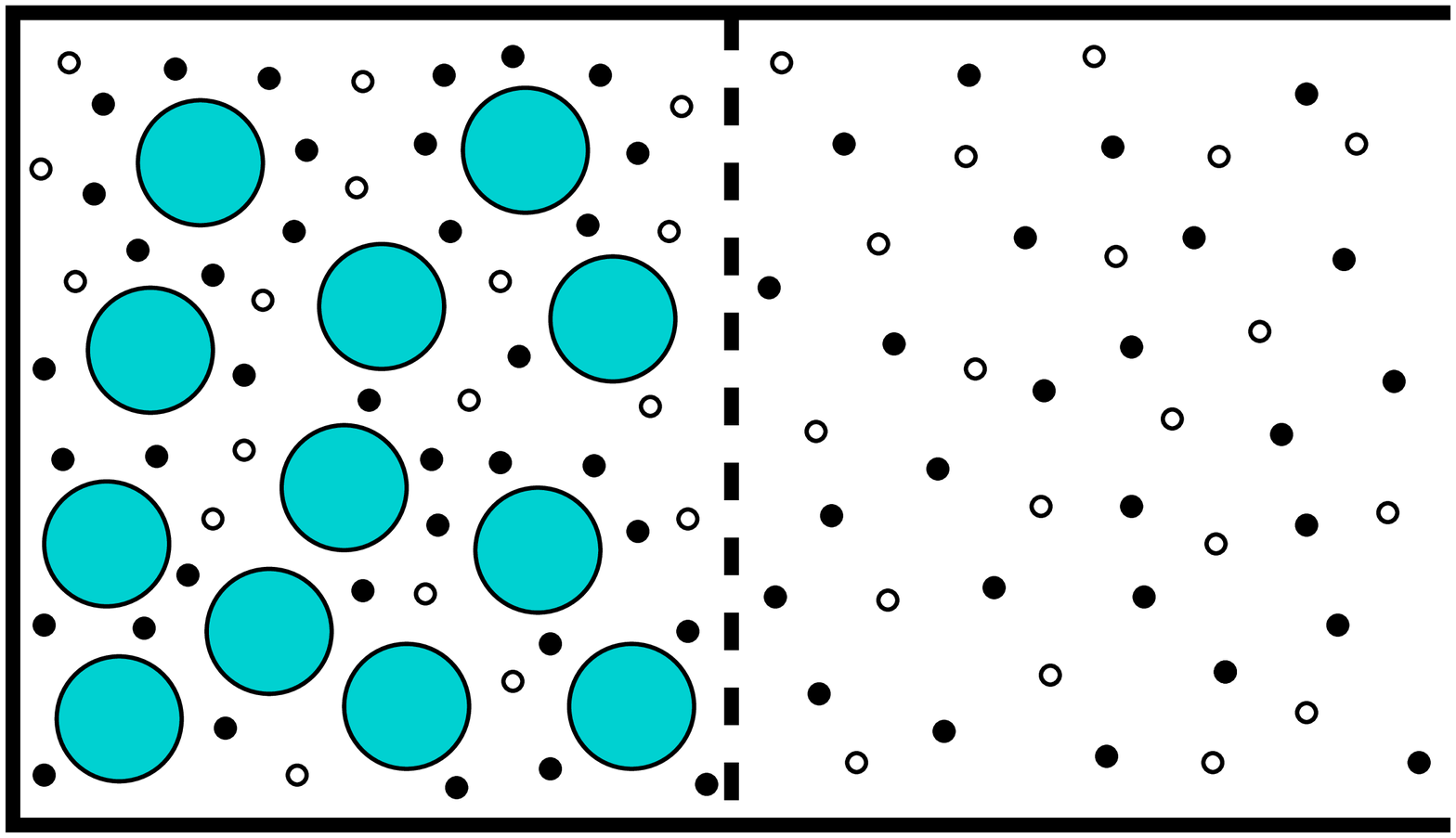}
\caption{\label{fig-model}
Primitive model of charged colloids: spherical macroions (radius $a$, valence $Z$)
and point microions dispersed in a dielectric continuum (left).
Colloidal suspension in Donnan equilibrium across semi-permeable membrane with 
electrolyte reservoir (right).
}
\end{center}
\end{figure}

If sufficiently dilute, the reservoir can be reasonably modeled as an ideal gas.
The reservoir fixes the chemical potential of salt in the suspension,
$\mu_s=2k_BT\ln(n_0\Lambda^3)$, where $n_0$ is the number density of ion pairs, 
$T$ the absolute temperature, and $\Lambda$ the thermal wavelength,
defining the arbitrary zero of the chemical potential.
A bulk suspension of $N_m$ macroions, $N_c$ counterions, and $N_s$ dissociated pairs
of oppositely charged salt ions contains $N_+=N_c+N_s$ positive and $N_-=N_s$
negative microions, for a total of $N_{\mu}=N_c+2N_s$ microions.
Global electroneutrality constrains the macroion and counterion numbers by the
condition $ZN_m=zN_c$.  

Accounting for excluded-volume and electrostatic (Coulomb) pairwise interparticle 
interactions, the Hamiltonian separates naturally into three terms: 
\begin{equation}
H=H_m+H_{\mu}+H_{m\mu}~.
\label{H}
\end{equation}
The first term on the right side is the macroion Hamiltonian
\begin{equation}
H_m=H_{\rm hs}+\frac{1}{2}\sum_{{i\neq j=1}}^{N_m}v_{mm}(r_{ij})~,
\label{Hm}
\end{equation}
where $H_{\rm hs}$ is the hard-sphere Hamiltonian (including kinetic energy),
$v_{mm}(r_{ij})=Z^2\lambda_B/r_{ij}$ is the Coulomb pair potential between
a pair of macroions with center-center separation $r_{ij}$, and 
$\lambda_B=e^2/(\epsilon k_BT)$ defines the Bjerrum length, the distance
at which the Coulomb interaction energy between a pair of monovalent ions
rivals the thermal energy.
To simplify notation, in this chapter, Hamiltonians, pairpotentials, and 
all other quantities having dimensions of energy are expressed in thermal ($k_BT$) units.
The remaining two terms on the right side of Eq.~(\ref{H}) are the microion Hamiltonian
\begin{equation}
H_{\mu}=K_{\mu}+\frac{\lambda_B}{2}\sum_{{i\neq j=1}}^{N_{\mu}}\frac{z_iz_j}{r_{ij}}~,
\label{Hmu}
\end{equation}
with kinetic energy $K_{\mu}$ and microion valences $z_i=\pm 1$, 
and the macroion-microion interaction energy
\begin{equation}
H_{m\mu}=Z\lambda_B\sum_{i=1}^{N_m}\sum_{j=1}^{N_{\mu}}\frac{z_j}{r_{ij}}~.
\label{Hmmu1}
\end{equation}

\section{One-Component Model: Effective Hamiltonian}\label{ocm}
Significant concentrations of salt pose severe computational challenges for 
large-scale simulations of bulk suspensions~\cite{linse00,brukhno09,hynninen07}.  
Consider that a suspension of macroions of radius $a=10$ nm at 1\% 
volume fraction and just 1~mM salt concentration contains, {\it per macroion},
${\cal O}(10^3)$ particles, all interacting via long-range Coulomb forces.
Therefore, salt-dominated suspensions usually are modeled by first mapping the
mixture onto a one-component model (OCM).

For a suspension in Donnan equilibrium, this mapping (or coarse graining)
operates on the semigrand partition function:
\begin{equation}
{\cal Z}=\langle\langle\exp(-H)\rangle_{\mu}\rangle_m~,
\label{Z1}
\end{equation}
where $\la~\ra_{\mu}$ denotes a grand canonical trace over microion coordinates
and $\la~\ra_m$ a canonical trace over macroion coordinates. 
The partition function may be formally expressed in the form
\begin{equation}
{\cal Z}=\la\exp(-H_{\rm eff})\ra_m~,
\label{Z2}
\end{equation}
where 
\begin{equation}
H_{\rm eff}=H_m+\Omega_{\mu}
\label{Heff}
\end{equation}
is an effective Hamiltonian of a one-component system of pseudo-macroions and
\begin{equation}
\Omega_{\mu}=-\ln\langle\exp(-H_{\mu}-H_{m\mu})\rangle_{\mu}
\label{Omegamu1}
\end{equation}
is the grand potential functional of the microions in the presence of fixed macroions.
Practical applications of the OCM require approximating $\Omega_{\mu}$.  For this purpose,
Poisson-Boltzmann (PB) theory~\cite{evans99,israelachvili92,deserno-holm01,denton10} 
is a powerful approach.  Below we briefly review PB theory and two common implementations: 
the cell model and the effective-interaction model.

\section{Poisson-Boltzmann Theory}\label{dft}
Poisson-Boltzmann theory is most elegantly formulated within the framework 
of classical density-functional theory of nonuniform 
fluids~\cite{dft-evans92,dft-evans79,dft-oxtoby91,lowen92,lowen_jcp93}.  Corresponding to
the primitive model Hamiltonian [Eqs.~(\ref{Hm})-(\ref{Hmmu1})], there exists a Helmholtz
free energy functional $F[n_m({\bf r}), n_{\pm}({\bf r})]$, which (for a given external
potential) is a unique functional of the macroion and microion number density profiles,
$n_m({\bf r})$ and $n_{\pm}({\bf r})$, varying with position ${\bf r}$~\cite{dft-evans79}.
This free energy functional separates, according to $F=F_{\rm id}+F_{\rm ex}+F_{\rm ext}$,
into a (purely entropic) ideal-gas free energy functional of all ions,
\begin{equation}
F_{\rm id}=\int{\rm d}{\bf r}\,\sum_{i=m,\pm}n_i({\bf r})\{\ln[n_i({\bf r})\Lambda^3]-1\}~,
\label{Fid}
\end{equation}
an excess free energy functional, $F_{\rm ex}=F_{\rm hs}+F_{\rm el}$, due to hard-sphere (hs)
and electrostatic (el) interparticle interactions, and a contribution $F_{\rm ext}$ due to
an external potential.  Neglecting interparticle correlations (mean-field approximation),
the electrostatic part of the excess free energy functional may be approximated as
\begin{equation}
F_{\rm el}=\frac{1}{e}\int{\rm d}{\bf r}\,\rho({\bf r})\Psi({\bf r})~,
\label{Fel}
\end{equation}
where $\rho({\bf r})=e[n_+({\bf r})-n_-({\bf r})-n_f({\bf r})]$ is the total
charge density, including the negative charge fixed on the macroion surfaces
of number density $n_f({\bf r})$, and
\begin{equation}
\Psi({\bf r})=\frac{\lambda_B}{e}\int{\rm d}{\bf r}'\,
\frac{\rho({\bf r}')}{|{\bf r}-{\bf r}'|}
\label{Phi}
\end{equation}
is the reduced electrostatic potential at position ${\bf r}$ due to all ions.
The reduced potential and total ion density are related via the Poisson equation,
which may be expressed in the form
\begin{equation}
\nabla^2\psi({\bf r})=-4\pi\lambda_B[n_+({\bf r})-n_-({\bf r})]~; \quad
\nabla\psi|_{\rm surface}=Z\lambda_B/a^2~,
\label{Poisson2}
\end{equation}
where the macroion charges are absorbed into a boundary condition at the 
macroion surfaces and the microion densities implicitly vanish inside the macroion cores.

In a given external potential, the equilibrium densities of all ions minimize the
total grand potential functional of the system~\cite{dft-evans79}.
Alternatively, fixing the macroions and regarding their charges as the source
of the external potential, the equilibrium microion densities alone minimize 
the {\it microion grand potential functional}
\begin{equation}
\Omega_{\mu}[n_{\pm}({\bf r})]=F_{\mu}[n_{\pm}({\bf r})]-
\mu_+\int{\rm d}{\bf r}\,n_+({\bf r})-\mu_-\int{\rm d}{\bf r}\,n_-({\bf r})~,
\label{Omega1}
\end{equation}
a Legendre transform of the microion free energy functional
\begin{equation}
F_{\mu}[n_{\pm}({\bf r})]=F_{\mu,{\rm id}}[n_{\pm}({\bf r})]+
\frac{1}{2e}\int{\rm d}{\bf r}\,\rho({\bf r})\Psi({\bf r})~,
\label{Fmu1}
\end{equation}
where 
\begin{equation}
F_{\mu,{\rm id}}[n_{\pm}({\bf r})]=
\int{\rm d}{\bf r}\,\sum_{i=\pm}n_i({\bf r})\{\ln[n_i({\bf r})\Lambda^3]-1\}
\label{Fmuid}
\end{equation}
is the ideal-gas free energy functional and the microion (electro)chemical potentials 
$\mu_{\pm}$ are identified as the Legendre variables.  Note that $\Omega_{\mu}$ 
depends parametrically on the macroion coordinates and that $F_{\mu}$ includes 
macroion-macroion Coulomb interactions for electroneutrality.  Under the assumption 
that either the electrostatic potential or the electric field vanishes everywhere 
on the boundary of the volume $V$, the microion free energy functional also may be
expressed in the form
\begin{equation}
F_{\mu}[n_{\pm}({\bf r})]=\int{\rm d}{\bf r}\,
\sum_{i=\pm}n_i({\bf r})\{\ln[n_i({\bf r})\Lambda^3]-1\}
+\frac{1}{8\pi\lambda_B}\int{\rm d}{\bf r}\,|\nabla\psi|^2~.
\label{FPB}
\end{equation}
Minimizing $\Omega_{\mu}[n_{\pm}({\bf r})]$ with respect to $n_{\pm}({\bf r})$
now yields the Boltzmann approximation for the equilibrium microion densities
\begin{equation}
n_{\pm}({\bf r})=n_{\pm}^{(0)}\exp[\mp\psi({\bf r})] \quad ({\rm fixed~macroions})~,
\label{boltzmann}
\end{equation}
where the reference densities, $n_{\pm}^{(0)}=\Lambda^{-3}\exp(\mu_{\pm})$, are the
microion densities at the reference potential $\psi=0$.  The microion grand potential
is the value of the grand potential functional [Eq.~(\ref{Omega1})] evaluated at the 
equilibrium density profiles [Eq.~(\ref{boltzmann})]:
\begin{equation}
\Omega_{\mu}=-\int{\rm d}{\bf r}\,[n_+({\bf r})+n_-({\bf r})]
-\frac{1}{2}\int{\rm d}{\bf r}\,[n_+({\bf r})-n_-({\bf r})
+n_f({\bf r})]\psi({\bf r})~.
\label{Omegaeq}
\end{equation}

For a closed suspension (fixed particle numbers), the chemical potentials of the two
microion species differ because of asymmetric interactions with the macroions:
$\mu_+\neq\mu_-$.  Correspondingly, the reference densities also differ:
$n_+^{(0)}\neq~n_-^{(0)}$.
In Donnan equilibrium, however, exchange of microions with a salt reservoir shifts the
intrinsic microion chemical potentials,
$\mu_{\pm}^{\rm in}=[\delta F_{\mu}/\delta n_{\pm}({\bf r})]_{\rm eq}$,
by the Donnan potential $\psi_D$:
\begin{equation}
\mu_{\pm}=\mu_{\pm}^{\rm in}\pm\psi_D=\ln(n_0\Lambda^3)~.
\label{mur}
\end{equation}
The total chemical potentials, and so too the reference densities, of the two microion
species are thus equalized.  The equilibrium microion density profiles are then given by
\begin{equation}
n_{\pm}({\bf r})=n_0\exp[\mp\psi({\bf r})]~.
\label{boltzmann-donnan}
\end{equation}
The Donnan potential is interpreted physically as the change in electrostatic potential
across the reservoir-suspension interface, and mathematically as a Lagrange multiplier
for the constraint of global electroneutrality.

Combining the Poisson equation for the potential [Eq.~(\ref{Poisson2})] with the
Boltzmann approximation for the microion densities [Eq.~(\ref{boltzmann})
or (\ref{boltzmann-donnan})], the Poisson-Boltzmann equation becomes
\begin{equation}
\nabla^2\psi({\bf r})~=~
\kappa_0^2\sinh\psi({\bf r})~; \quad \nabla\psi|_{\rm surface}=Z\lambda_B/a^2~,
\label{PB2}
\end{equation}
where $\kappa_0=\sqrt{8\pi\lambda_B n_0}$ is the screening constant in the reservoir.
Note that Eq.~(\ref{PB2}) is highly nonlinear.
In the case of weak electrostatic potentials ($\psi\ll 1$), the right side of 
Eq.~(\ref{PB2}) may be expanded in powers of $\psi$.  Truncating at linear order
yields the linearized PB equation:
\begin{equation}
\nabla^2\psi({\bf r})~=~
\kappa_0^2\psi({\bf r})~; \quad \nabla\psi|_{\rm surface}=Z\lambda_B/a^2~.
\label{PBlin}
\end{equation}
Beyond the boundary condition at the macroion surfaces, the boundary-value problem is
fully specified only by imposing another condition at the outer boundary of the system,
which depends on the practical implementation of the theory.

\section{Cell-Model Implementation}\label{cm}
\begin{figure}[h!]
\begin{center}
\hspace*{1cm}
\includegraphics[width=0.4\columnwidth]{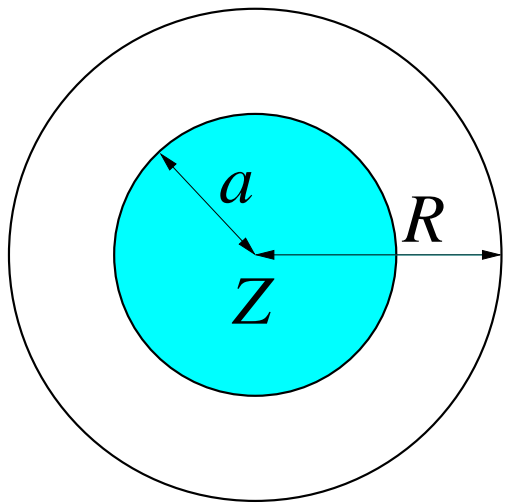}
\caption{\label{pbcell}
Cell model: a single macroion in a spherical cell.
}
\end{center}
\end{figure}
The anisotropic boundary conditions on the nonlinear PB equation [Eq.~(\ref{PB2})]
imposed by an arbitrary configuration of macroions render a general solution 
computationally daunting.  In recent years, powerful {\it ab initio} methods
have been developed for combining PB theory of microion density profiles with
molecular dynamics~\cite{lowen92,lowen_jcp93,lowen_el93,tehver99} or Brownian
dynamics~\cite{dobnikar03,dobnikar04} simulation to evolve macroion coordinates
according to derived forces.  Despite such advances, most applications of PB theory 
have been implemented within a cell model to facilitate numerical solution.

In a seemingly bold reduction, the cell model represents a bulk suspension by a
single macroion, neutralizing counterions, and salt ions confined to a cell
of the same shape as the macroion (see Fig.~\ref{pbcell}).  Microion-induced 
correlations between macroions are simply ignored.  For spherical colloids, 
the natural choice is a spherical cell centered on the macroion.  Gauss's law 
then dictates that the electric field must vanish everywhere on the boundary of 
the electroneutral cell.
With the potential and microion densities depending on only the radial coordinate
$r$, the PB equation reduces to an ordinary differential equation for $\psi(r)$
with boundary conditions $\psi'(a)=Z\lambda_B/a^2$ and $\psi'(R)=0$,
where the cell radius $R$ is commensurate with the average macroion density:
$n_m=N_m/V=3/(4\pi R^3)$.  For a closed suspension, the arbitrary location
of the reference point of the electrostatic potential (where $\psi=0$) is usually
chosen as the cell boundary: $\psi(R)=0$.  In Donnan equilibrium, the potential
is conventionally chosen to vanish in the reservoir, in which case the boundary value
of the electrostatic potential is identified as the Donnan potential:
$\psi(R)=\psi_D\neq~0$.

An appealing feature of the cell model is the simple analytic relation between the
bulk pressure $p$ (in thermal units) and the microion densities at the cell boundary:
\begin{equation}
p=n_+(R)+n_-(R)~.
\label{pressure}
\end{equation}
Although first derived within the mean-field PB framework~\cite{marcus55},
this pressure theorem proves to be exact within the
cell model~\cite{wennerstrom82}, i.e., valid also for correlated microions.
In Donnan equilibrium with an ideal-gas reservoir, the osmotic pressure $\Pi$,
i.e., the difference in pressure between suspension and reservoir,
is then given by
\begin{equation}
\Pi=n_+(R)+n_-(R)-2n_0~.
\label{osmotic-pressure}
\end{equation}
Within PB theory, the osmotic pressure is strictly positive~\cite{deserno02},
following from Eqs.~(\ref{boltzmann-donnan}) and (\ref{osmotic-pressure}) 
and the inequality, $\cosh x>1$.

The cell model provides one means of implementing PB theory by approximating the
microion grand potential [Eq.~(\ref{Heff})] in the one-component mapping of the
primitive model (Sec.~\ref{ocm}).  Reducing a suspension to a single macroion in a
cell with isotropic boundary conditions facilitates solution of the nonlinear PB equation.
The cost of incorporating nonlinear microion screening, however, is neglect 
of correlations between macroions.

\section{Effective-Interaction Implementation}\label{effint}
An alternative implementation of PB theory, based on the one-component mapping,
focuses on effective interactions derived from perturbative expansion of the microion
grand potential about a reference system, namely, a uniform plasma of microions
unperturbed by the macroions.  By incorporating macroion interactions, this approach
can model both thermodynamic and structural properties of colloidal suspensions.
For reviews of effective interactions in colloidal suspensions, see 
refs.~\cite{belloni00,hansen-lowen,likos01,zvelindovsky07}.
Several statistical mechanical frameworks have been developed.
Density-functional theories~\cite{vRH97,graf98,vRDH99,zoetekouw_prl06} expand 
the ideal-gas part of the microion free energy functional [Eq.~(\ref{Fmuid})] 
in a Taylor series in powers of deviations of the microion density profiles 
from their mean values $n_{\pm}$:
\begin{equation}
F_{\mu,{\rm id}}[n_{\pm}({\bf r})]=\sum_{i=\pm}\left(N_i[\ln(n_i\Lambda^3)-1]
+\int{\rm d}{\bf r}\,\frac{[n_i({\bf r})-n_i]^2}{2n_i}+\cdots\right)~.
\label{Fid-DFT}
\end{equation}
Distributon function theories are based on extensions of the Debye-H\"uckel 
theory of electrolytes~\cite{warren00}.
Response theories~\cite{silbert91,denton99,denton00,denton07} are based on a 
similar perturbative expansion of the microion grand potential functional 
[Eq.~(\ref{Omegamu1})]:
\begin{equation}
\Omega_{\mu}=\Omega_0+\int_0^1 d\lambda\, \langle H_{m\mu}\rangle_{\lambda}~,
\label{Omegamu2}
\end{equation}
where $\Omega_0=-\ln\langle\exp(-H_{\mu})\rangle_{\mu}$ is the grand potential 
of a reference suspension in the absence of an external colloidal potential and 
$\langle H_{m\mu}\rangle_{\lambda}$ denotes an ensemble average of the 
macroion-microion interaction energy in a system in which the macroions are 
``charged" to a fraction $\lambda$ of their full charge, which can be related to 
the macroion-microion pair potentials $v_{m\pm}(r)={\pm}Z\lambda_B/r$ and the 
densities, $n_m({\bf r})$ and $n_{\pm}({\bf r})$, of macroions and microions:
\begin{equation}
\la H_{m\pm}\ra_{\lambda}=\int d{\bf r}\, \int d{\bf r}'\, 
v_{m\pm}(|{\bf r}-{\bf r}'|) n_m({\bf r}) \la n_{\pm}({\bf r}')\ra_{\lambda}~.
\label{Hmmu2}
\end{equation}
Expanding the microion densities about a reference microion plasma 
in powers of the macroion ``external" potential,
\begin{equation}
\phi_{\pm}({\bf r})=\int{\rm d}{\bf r}'\,v_{m\pm}(|{\bf r}-{\bf r}'|)n_m({\bf r}')~,
\label{phipm}
\end{equation}
and truncating the series at linear order yields the linear-response approximation
\begin{equation}
n_{\pm}({\bf r})=\sum_{i=\pm}\int{\rm d}{\bf r}'\,\chi_{\pm i}(|{\bf r}-{\bf r}'|)
\phi_i({\bf r}')~.
\label{npmr1}
\end{equation}
The linear-response functions
\begin{equation}
\chi_{ij}(|{\bf r}-{\bf r}'|)=\left(\frac{\delta n_i({\bf r})}{\delta\phi_j({\bf r}')}\right)_{Z=0}
\label{chir1}
\end{equation}
describe the response of the reference plasma to the macroions and are 
related to the plasma pair correlation functions $h_{ij}(r)$ via~\cite{HM}
\begin{equation}
\chi_{ij}(r)
=-n_i[\delta_{ij}\delta(r)+n_j h_{ij}(r)]~.
\label{chir2}
\end{equation}
The neglected higher-order terms in Eq.~(\ref{npmr1}) involve 
nonlinear response functions and many-particle correlations.

Approximating the pair correlation functions of the uniform plasma by their 
asymptotic (long-range) limits yields the so-called random phase approximation 
\begin{equation}
h_{ij}(r)
=-z_iz_j\lambda_B\frac{e^{-\kappa r}}{r}~,
\label{rpa}
\end{equation}
where $\kappa=\sqrt{4\pi\lambda_B n_{\mu}}$ is the screening constant in the system,
which differs from that in the reservoir $\kappa_0$ [{\it cf}.~Eq.~(\ref{PB2})].
For consistency, $n_{\mu}$ here represents the total density of microions in
the {\it free} volume, i.e., the volume not excluded by the macroion hard cores.

Inserting the linearized microion densities [Eq.~(\ref{npmr1})] into Eq.~(\ref{Hmmu2}) 
recasts the effective Hamiltonian (microion grand potential) in the form of a sum of 
effective interactions:
\begin{equation}
H_{\rm eff}=E_v+H_{\rm hs}+\frac{1}{2}\sum^{N_m}_{i\neq j=1}v_{\rm eff}(r_{ij})~,
\label{Heff2}
\end{equation}
where the volume energy $E_v$ is the microion grand potential for a single macroion,
$v_{\rm eff}(r)$ is an {\it effective} electrostatic pair potential between macroions,
and neglected higher-order terms involve sums over effective many-body interactions.  
The volume energy takes the general form
\begin{equation}
E_v=\Omega_0+\frac{1}{2}N_mv_{\rm ind}(0)+\frac{1}{2}(N_+-N_-)\Psi_D~,
\label{Ev}
\end{equation}
where the first term on the right side accounts for the microion entropy,
the second term for the macroion-microion interaction energy, and the last term
is the Donnan potential energy resulting from the electroneutrality constraint.
The effective macroion-macroion pair potential can be expressed as 
\begin{equation}
v_{\rm eff}(r)=v_{mm}(r)+v_{\rm ind}(r)~,
\label{veffr1}
\end{equation}
which comprises the bare Coulomb pair potential $v_{mm}(r)$ and a 
{\it microion-induced} pair potential
\begin{equation}
v_{\rm ind}(r)= 
\int{\rm d}{\bf r}'\, [n_+(r')-n_-(r')]v_{m+}(|{\bf r}-{\bf r}'|)~.
\label{vindr1}
\end{equation}
To evaluate $v_{\rm ind}(r)$, it is helpful first to consider the isotropic
microion density profiles around a single macroion, which are obtained 
from Eqs.~(\ref{npmr1})-(\ref{rpa}):
\begin{eqnarray}
n_{\pm}(r)&=&
\sum_{i=\pm}\int{\rm d}{\bf r}'\,\chi_{\pm i}(|{\bf r}-{\bf r}'|)
v_{m+}(r') 
\nonumber\\
&=&\pm n_{\pm}
\left(
n_{\mu}\lambda_B\int d{\bf r}'\, 
\frac{e^{-\kappa r'}}{r'}v_{m+}(|{\bf r}-{\bf r}'|)
-v_{m+}(r)
\right).
\qquad\quad
\label{npmr2}
\end{eqnarray}
To ensure exclusion of microions from the macroion hard cores, the 
macroion-microion pair potentials can be extended inside the core:
\begin{equation}
v_{m\pm}(r)=\pm Z\lambda_B f(r); \qquad f(r)=
\left\{ \begin{array}
{l@{\quad\quad}l}
1/r~, 
& r>a \\[2ex]
\alpha/a~, 
& r<a~, \end{array} \right.
\label{vmpmr}
\end{equation}
where $\alpha$ is a constant to be chosen such that $n_{\pm}(r)=0$ for $r<a$.
After evaluating the convolution integral
\begin{equation}
\int_0^{\infty} dr'\, r'e^{-\kappa r'}g(r,r')=
\frac{2}{\kappa^2}\left\{ \begin{array}
{l@{\quad\quad}l}
\frac{\displaystyle 1}{\displaystyle r}
-\frac{\displaystyle e^{\kappa a}}{\displaystyle 1+\kappa a} 
\frac{\displaystyle e^{-\kappa r}}{\displaystyle r}~, 
& r>a \\[2ex]
\frac{\displaystyle 1}{\displaystyle a+\kappa^{-1}}~, 
& r<a~, \end{array} \right.
\label{nasty-integral1}
\end{equation}
with
\begin{equation}
g(r,r')=\int_{-1}^1d\mu\, f(|{\bf r}-{\bf r}'|)~; \quad
\mu\equiv{\bf r}\cdot{\bf r}'/(rr')~,
\label{g}
\end{equation}
one finds [from Eqs.~(\ref{npmr2}) and (\ref{vmpmr})] $\alpha=\kappa a/(1+\kappa a)$ and
\begin{equation}
n_{\pm}(r)=\mp n_{\pm}Z\lambda_B\left(\frac{e^{\kappa a}}{1+\kappa a}\right)
\frac{e^{-\kappa r}}{r}~,
\qquad r>a~.
\label{npmr3}
\end{equation}
Substituting Eq.~(\ref{npmr3}) into Eqs.~(\ref{veffr1}) and (\ref{vindr1}),
and evaluating the integral (for $r>2a$)
\begin{equation}
\int_a^{\infty}dr'\, r'e^{-\kappa r'}g(r,r')= 
\frac{2}{\kappa^2 r}
\left(
\frac{1+\kappa a}{e^{\kappa a}}
-\frac{e^{\kappa a}}{1+\kappa a}e^{-\kappa r}
\right)~,
\label{nasty-integral2}
\end{equation}
yields a screened-Coulomb (Yukawa) effective pair potential:
\begin{equation}
v_{\rm eff}(r)= 
Z^2\lambda_B\left(\frac{e^{\kappa a}}{1+\kappa a}\right)^2
\frac{e^{-\kappa r}}{r}~,
\qquad r>2a~.
\label{veffr2}
\end{equation}
Finally, from Eqs.~(\ref{vindr1}) and (\ref{npmr3}), we have
\begin{equation}
v_{\rm ind}(0)=-\frac{Z^2\lambda_B}{a+\kappa^{-1}}~,
\label{vind0}
\end{equation}
which allows the volume energy to be written more explicitly:
\begin{equation}
E_v=\Omega_0-\frac{1}{2}N_m\frac{Z^2\lambda_B}{a+\kappa^{-1}}
-\frac{1}{2}\frac{(N_+-N_-)^2}{N_{\mu}}~.
\label{Ev2}
\end{equation}
In passing, we recall that Eq.~(\ref{veffr2}) is the basis of the classic 
Derjaguin-Landau-Verwey-Overbeek (DLVO) theory~\cite{DL,VO} of charged colloids,
where it was first derived within a Debye-H\"uckel approximation, without the 
associated volume energy and without excluded-volume corrections.

It is important to remember that the one-body volume energy, although
independent of macroion coordinates, does depend on the {\it average} 
macroion density, and therefore can influence thermodynamic properties.
Similarly, the effective pair potential is density-dependent.  
Implications of density-dependent effective interactions for thermodynamic 
stability and consistency have been discussed extensively in recent 
years~\cite{louis02,dobnikar06,castaneda-priego06,trizac07,schurtenberger08,castaneda-priego12}.

Bulk thermodynamic and structural properties can be calculated ultimately by 
inputting the effective interactions, $E_v$ and $v_{\rm eff}(r)$, 
into statistical mechanical theories or simulations of the OCM.
Figures~\ref{pressure-linse} and \ref{pressure-jonsson} show sample results 
for the osmotic pressure of highly charged colloidal suspensions, 
calculated using methodologies described in refs.~\cite{denton10,lu-denton10,denton08}.  
Predictions of the PB cell and effective interaction (one-component) models 
agree closely with simulations of the primitive model~\cite{linse00} 
(Fig.~\ref{pressure-linse}) up to moderate electrostatic coupling strengths 
($\Gamma\equiv\lambda_B/a<1$) and with experimental data~\cite{jonsson11,chang95} 
(Fig.~\ref{pressure-jonsson}) over a considerable range of colloid volume fractions.
To achieve such agreement, the linear-response theory has been merged with 
charge renormalization theory to incorporate the important concept of an 
effective macroion valence~\cite{alexander84}.  By subsuming within the 
effective valence those counterions that are strongly associated with the macroions,
the renormalized theory includes most of the nonlinear response inherent in 
the PB equation.  Comparably accurate results are obtained with a 
renormalized jellium theory~\cite{levin07,levin09,castaneda-priego11}.
The effective interaction model also predicts 
radial distribution functions in close agreement with simulations of the 
primitive model~\cite{castaneda-priego12,lu-denton10}.
\begin{figure}[t!]
\begin{center}
\vspace*{0.5cm}
\includegraphics[width=0.7\columnwidth]{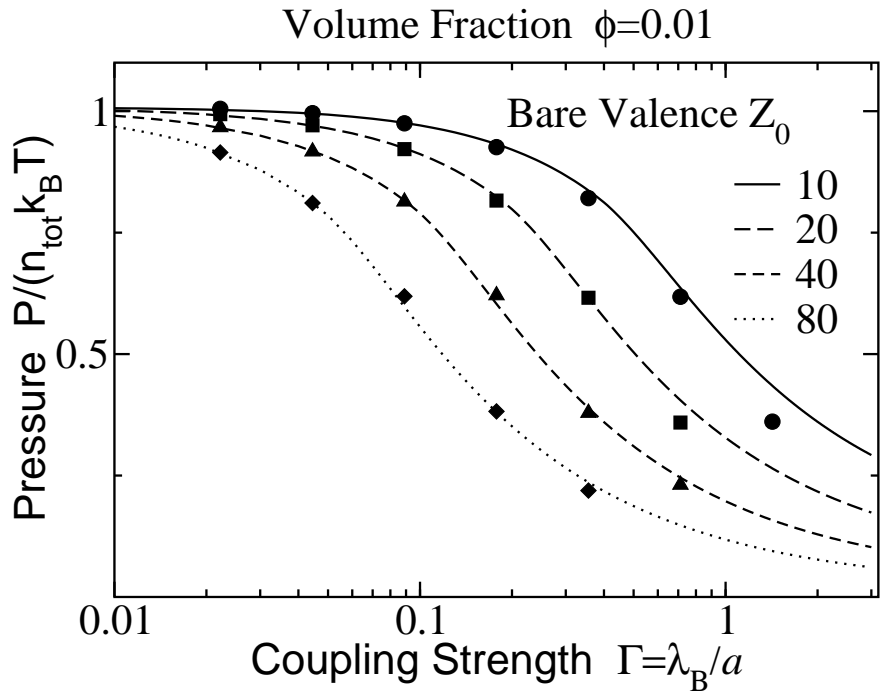}
\caption{\label{pressure-linse}
Pressure vs. electrostatic coupling constant $\Gamma\equiv\lambda_B/a$
for salt-free suspensions with colloid volume fraction $\phi=0.01$ and
several bare valences $Z_0$.
Predictions of the effective interaction model (curves) are compared with 
primitive model simulation data (symbols) [7].
}
\end{center}
\end{figure}
\begin{figure}
\begin{center}
\vspace*{0.5cm}
\includegraphics[width=0.7\columnwidth]{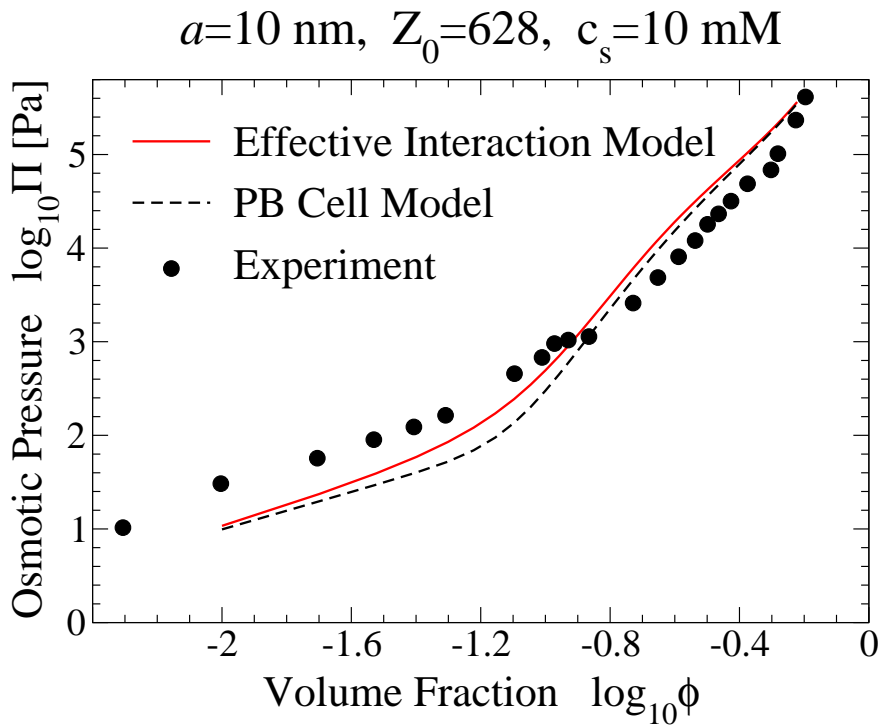}
\caption{\label{pressure-jonsson}
Osmotic pressure vs. colloid volume fraction for suspensions with
macroion radius $a=10$ nm, bare valence $Z_0=628$, and salt concentration $c_s=10$ mM.
Predictions of the effective interaction model (solid curve) and the
PB cell model (dashed curve) are compared with experimental data (symbols) [44, 45].
}
\end{center}
\end{figure}

\section{Outlook}\label{outlook}
To illustrate the essence of coarse-grained modeling in a friendly context,
the discussion in this chapter is limited to the primitive model of charged colloids,
focusing on monodisperse suspensions of microspheres.  For this system, the effective 
electrostatic interactions derived using the coarse-graining scheme described in 
Sec.~\ref{effint} are relatively simple and lead to predictions for thermodynamic 
and structural properties that agree closely with the Poisson-Boltzmann cell model,
detailed simulations of the primitive model, and experiment.
Tremendous opportunities now lie ahead for extending the general methods outlined 
here and applying them to more complex systems, such as mixtures of colloids 
differing in size, shape, and charge, to suspensions of particles with anisotropic 
(patchy) charge distributions (e.g., Janus particles), as well as to different 
interparticle interactions, such as dipolar interactions.
As computing power grows, the demand for coarse-grained models will likely persist,
to guide and interpret simulations and to facilitate exploration of increasingly
complex materials.

\ack
It is a pleasure to thank Jun Kyung Chung and Sylvio May for helpful discussions.
This work was supported by the National Science Foundation
under Grant No.~DMR-1106331.



\end{document}